\documentstyle[prb,aps,preprint,graphicx]{revtex}

\tighten
\begin{document}

\title{Exact Spectral Functions of a Non Fermi Liquid in 1 Dimension   }

\author{ Karlo Penc$^{1}$ and B. Sriram Shastry$^2$}

\address{$^1$ Research Institute for Solid State Physics and Optics, 
H-1525
  Budapest, P.O.B. 49, Hungary \\
$^2$ Bell Laboratories, Lucent Technologies\\
Murray Hill , NJ, and \\
IISc, Bangalore 560012 , India 
 }

\date{\today}
\maketitle

\begin{abstract}
We study the exact one electron propagator and spectral function of 
 a solvable model of interacting electrons due to Schulz and Shastry. 
The solution  previously found for the energies and wave functions 
is extended to give the spectral functions, which turn out to be 
computable, interesting and non trivial. They provide one of the 
few examples of cases where the spectral functions are known 
asymptotically as well as exactly. 
\end{abstract}
 \narrowtext
\section{Introduction}
Schulz and Shastry \cite{SS} have introduced a new class of gauge coupled
one-dimensional (1D) Fermi systems that are non Fermi liquid in the sense that
the momentum distribution function has a cusp at the Fermi momentum $k_{F}$
rather than a jump as in a Fermi liquid.\cite{kfcusp} This behavior is of the
sort first found by Luttinger in the context of his study of a one dimensional
model that is popularly known as the Luttinger model.\cite{lutmod} The model
introduced by Schulz and Shastry (SS) is in fact intimately connected to the
Luttinger model, and is best viewed as a reinterpretation of Luttinger's
original model as a gauge theory. Particles of different species exert a
mutual gauge potential on each other, and this is sufficient to destroy the
Fermi liquid. This model has the added property that the charge and spin
correlations are unaffected by the interaction, owing to the ``gauge'' nature
of interaction. The fermionic Green's functions, however, are non trivial, and
have characteristic singularities that are popularly known as the
Tomonaga-Luttinger liquid behavior.\cite{lutliq,voit}  The asymptotic long
distance behavior of the one electron correlation function is known (see
below) by one of several arguments, including Luttinger's original one.

Our motivation in the present work is to compute the exact one electron
Green's function for the SS\ model, utilizing our knowledge of the complete
spectrum of the same, and using techniques familiar from Anderson's treatment
of the Orthogonality Catastrophe issue in the X-ray edge 
problem.\cite{anderson67}  This is of great interest since usually one does not have
access to the exact Green's function even in 1D, and one has to be content
with the asymptotic behavior. For interpreting experiments, such as those on
photoemission, one wants to know more than just the asymptotics, and this
possibility is realized here for the particular model of SS.

We first write down the basic lattice Fermi model in 1D, outline the 
pseudo
unitary transformation that eliminates the gauge interactions in favor 
of a
twisted boundary condition. Using this transformation we formulate the
problem of calculating the one electron Green's function. 

\section{The model}
  Let us write the model for two component electrons hopping and 
interacting
via the Hamiltonian
\begin{equation}
  {\cal H} = -t \sum_{j=0}^{L-1} \sum_{\sigma} 
  \exp(i\sigma \alpha [\hat n_{j,\bar\sigma} \!+\! \hat 
n_{j+1,\bar\sigma}])
  c^\dagger_{j,\sigma} c^{\phantom{\dagger}}_{j+1,\sigma} + {\rm H.c.} \,,
 \label{eq:HSS}
\end{equation}
where for concreteness we have simplified the original model presented 
in Ref.~\onlinecite{SS}. The unitary transformation 
\begin{equation}
 {\cal U}_1 = \exp(i \sum_{l > m} \alpha [\hat n_{l,\uparrow} \hat 
n_{m,\downarrow}
                                -\hat n_{m,\uparrow} \hat 
n_{l,\downarrow}])
 \label{eq:U1}
\end{equation}
transforms the Eq.~(\ref{eq:HSS}) into a simple hopping Hamiltonian 
with 
twisted boundary conditions.\cite{SS} To regain a translational 
invariant
Hamiltonian we apply a second unitary transformation
\begin{equation}
 {\cal U}_2 = \prod_{l=0}^{L-1} \exp \frac{2i \alpha l (
 \hat N_{\uparrow} \hat n_{l,\downarrow}-
  \hat N_{\downarrow} \hat n_{l,\uparrow})}{L} \,.
 \label{eq:U2}
\end{equation}
 The combined transformation ${\cal U} ={\cal U}_2{\cal U}_1$ commutes
with ${\cal T}$, where ${\cal T}$ is the translational operator which 
shifts
one site to the right (e.g. ${\cal T} \hat n_{j} {\cal T}^\dagger = 
\hat n_{j+1}$).  
The effect of ${\cal U}$ on the fermion operators is
\begin{eqnarray}
{\cal U} c^\dagger_{j,\sigma} {\cal U}^\dagger 
&=& 
  e^{-i\alpha\sigma N_{\bar\sigma}}  
  e^{i\alpha\sigma \hat n_{j,\bar\sigma}}
  c^\dagger_{j,\sigma}
 \prod_{l=0}^{j-1} \exp(2i\alpha\sigma \hat n_{l,\bar\sigma})
 \prod_{l=0}^{L-1} 
  \exp \frac{2 i\sigma (l-j)\alpha  \hat n_{l,\bar \sigma}}{L}
 \nonumber \\
{\cal U} c^{\phantom{\dagger}}_{j,\sigma} {\cal U}^\dagger &=& 
  e^{i\alpha\sigma N_{\bar\sigma}}  
  e^{-i\alpha\sigma \hat n_{j,\bar\sigma}}
  c^{\phantom{\dagger}}_{j,\sigma}
 \prod_{l=0}^{j-1} \exp(-2i\alpha\sigma \hat n_{l,\bar\sigma})
 \prod_{l=0}^{L-1} 
  \exp \frac{-2 i\sigma(l-j)\alpha \hat n_{l,\bar \sigma}}{L} \,,
 \label{eq:UU}
\end{eqnarray}
while the density operators are invariant, 
${\cal U} \hat n_{j,\sigma} {\cal U}^\dagger = \hat n_{j,\sigma}$, 
and the transformed Hamiltonian 
$\widetilde {\cal H}={\cal U} {\cal H} {\cal U}^\dagger$
reads
\begin{equation}
\widetilde {\cal H} =  -t \sum_{j=0}^{L-1} \sum_{\sigma} 
     \left(e^{2i\sigma\alpha n_{\bar\sigma}}
c^\dagger_{j,\sigma} c^{\phantom{\dagger}}_{j+1,\sigma} + {\rm 
H.c.}\right) \,,
\end{equation}
where $n_\sigma=N_\sigma/L$ is the density of $\sigma$ spin fermions.
Thus we see that the transformed hopping has a ``dynamically 
generated'' gauge field. 
In the eigenvalue problem 
\begin{equation}
 \widetilde{\cal H}|\tilde \phi \rangle = E | \tilde \phi \rangle
\end{equation}
the eigenstates $|\tilde \phi \rangle$ 
are products of noninteracting one-particle states with 
momenta $k$ created with 
$c_{k,\sigma }^{\dagger}= L^{-1/2} \sum_{l} e^{ikl} c_{l,\sigma 
}^{\dagger}$ 
operator, 
$|\tilde \phi \rangle = \prod_{k,\sigma }c_{k\sigma }^{\dagger }|0 
\rangle$.
 The momenta are quantized as
$ L k_{j,\sigma} = 2 \pi {\cal I}_{j,\sigma}$ with ${\cal 
I}_{j,\sigma}$
integers.
The total energy and momentum of the states is 
\begin{equation}
  E =  \sum_\sigma \sum_{j=1}^{N_\sigma} 
\varepsilon_\sigma(k_{j,\sigma}),
  \quad 
  P = \sum_\sigma \sum_{j=1}^{N_\sigma} k_{j,\sigma} \,, \label{eq:EP} 
\end{equation}
and the one-particle energy is 
\begin{equation}
 \varepsilon_\sigma(k) = -2t \cos(k+2\sigma\alpha n_{\bar\sigma}) 
\label{eq:ene1p} \,.
\end{equation} 
Thus we must have the eigenstates of ${\cal H}$ 
\begin{equation}
  \label{eq:canpsi}
  | \phi \rangle= U^{\dagger }| \tilde \phi \rangle \,, 
\end{equation}
with the energy and momentum 
given also by Eq.~(\ref{eq:EP}). In the ground state the $k$ states 
between
the Fermi momenta 
$k_{F,\sigma}^{-}$ and $k_{F,\sigma}^{+}$ are filled 
($k_{F,\sigma}^{\pm}=\pm\pi n_\sigma - 2 \alpha \sigma n_{\bar\sigma}$).
 In the thermodynamic limit the energy $E$ does not depend on $\alpha$ 
 and is
 equal with the energy of the noninteracting $\alpha=0$ case. For finite 
 size
 systems $\alpha$ enters only through the $O(1/L)$ corrections. 

\section{Spectral functions}
Our goal is to calculate the spectral functions, which we define as
\begin{eqnarray}
  A_\sigma(k,\omega) &=& \sum_{f} 
  |\langle f | c^\dagger_{k,\sigma} | {\rm GS} \rangle|^2 
  \delta(\omega-E_f^{N+1}+E_{\rm GS}), \label{eq:akw}\\
  B_\sigma(k,\omega) &=& \sum_{f} 
  |\langle f | c^{\phantom{\dagger}}_{k,\sigma} | {\rm GS} \rangle|^2 
  \delta(\omega-E_{\rm GS}+E_f^{N-1}) \,. \label{eq:bkw}
\end{eqnarray}
The local ($k$ averaged) spectral functions are defined as 
\begin{eqnarray}
  A_\sigma(\omega) &=& \frac{1}{L} \sum_{k}   A_\sigma(k,\omega)\,, \\
B_\sigma(\omega) &=& \frac{1}{L} \sum_{k} B_\sigma(k,\omega)\,.
\end{eqnarray}
We concentrate on $A_\uparrow(k,\omega)$, since $B_\sigma(k,\omega)$ is 
calculated analogously.

As we mentioned in the introduction, 1D interacting fermions behave as
Luttinger-liquids, which is characterized, among others, by the power-law
behavior of correlation function for small energies.
In our case, as we will see later, the main contribution for $0<\alpha<\pi$
comes from
\begin{eqnarray}
  A_\uparrow(k,\omega) &\approx& 
  c_1 
  \frac{
    \left[
      (\omega-\varepsilon_F)^2-u^2(k-k^{(-1)}_{\uparrow})^2
    \right]^{(\alpha/\pi)^2}}
   {  \omega-\varepsilon_F-u(k-k^{(-1)}_{\uparrow})}
+  c_1 
  \frac{
     \left[
        (\omega-\varepsilon_F)^2-u^2(k-k^{(1)}_{\uparrow})^2
     \right]^{(\alpha/\pi)^2}}
  {  \omega-\varepsilon_F+u(k-k^{(1)}_{\uparrow})}
  \nonumber\\
  && +  c_2   
\frac{\left[(\omega-\varepsilon_F)^2-u^2(k-k^{(1)}_{\uparrow})^2\right]^{(\alpha/\pi-1)^2}}{  \omega-\varepsilon_F-u(k-k^{(1)}_{\uparrow})}
  + c_2 
  \frac{\left[
     (\omega-\varepsilon_F)^2-u^2(k-k^{(3)}_{\uparrow})^2
   \right]^{(\alpha/\pi-1)^2}}
   {\omega-\varepsilon_F+u(k-k^{(3)}_{\uparrow})}\,,
 \label{eq:powAkw}
\end{eqnarray}
where 
$k^{(\nu)}_\sigma= \nu \pi n_\sigma -  2 \sigma \alpha n_{-\sigma}$  
%$k^{(\nu)}_\uparrow= \nu \pi n_\uparrow - 2 \alpha n_{\downarrow}$ 
are the
(Fermi) momenta of the singularities, $c_1$ and $c_2$ are constants and $u$ is
the velocity of the excitations. Usually in Luttinger-liquids the
velocities of the spin and charge excitations are different 
and they both appear
in spectral functions.  In our case, however, due to the gauge origin of the
interaction, the spin and charge velocities are equal to the fermi
velocity $v_F$.  The spectral function has a nonanalytical, branch cut
structure near the fermi momenta, and finite weight appears for higher
multiples of the fermi momenta ($k^{(3)}_{\uparrow}$).
The local density of states near Fermi energy reads
\begin{equation}
A(\omega)\approx c_1 (\omega-\varepsilon_F)^{2(\alpha/\pi)^2} + 
  c_2 (\omega-\varepsilon_F)^{2(\alpha/\pi-1)^2} \,,
 \label{eq:aw_le}
\end{equation}
which for the noninteracting $\alpha=0$ reproduces the 
Fermi-liquid step function.

We now consider the exact evaluation of the spectral functions.
As a preliminary to  the discussion for general $\alpha$, let us 
note 
the special cases of $\alpha=0$ and $\alpha=\pi$, where the spectral 
functions
can be calculated more or less trivially.

(i) The  $\alpha=0$  case is nothing else but the usual tight binding 
Hamiltonian
\begin{equation}
  {\cal H} = -t \sum_{j,\sigma} 
     (c^\dagger_{j,\sigma} c^{\phantom{\dagger}}_{j+1,\sigma} + {\rm 
H.c.})
\end{equation}
of noninteracting electrons, as $e^{i \alpha \hat n_j}=1$ in 
Eq.~(\ref{eq:HSS}).
For
the spectral functions we recover the familiar
\begin{eqnarray}
 A^{(0)}_\sigma(k,\omega) &=& \delta(\omega+2 t \cos k) 
   \Theta(\omega - \varepsilon_F) \,, \label{eq:Akw0}
\\
 B^{(0)}_\sigma(k,\omega) &=& \delta(\omega+2 t \cos k) 
   \Theta(\varepsilon_F-\omega) \,, 
 \label{eq:Bkw0}
\end{eqnarray}
i.e. a Dirac-delta peak following the cosine-like dispersion of the free
fermions. 
  
\begin{figure}
\centering
\includegraphics[width=7.0truecm]{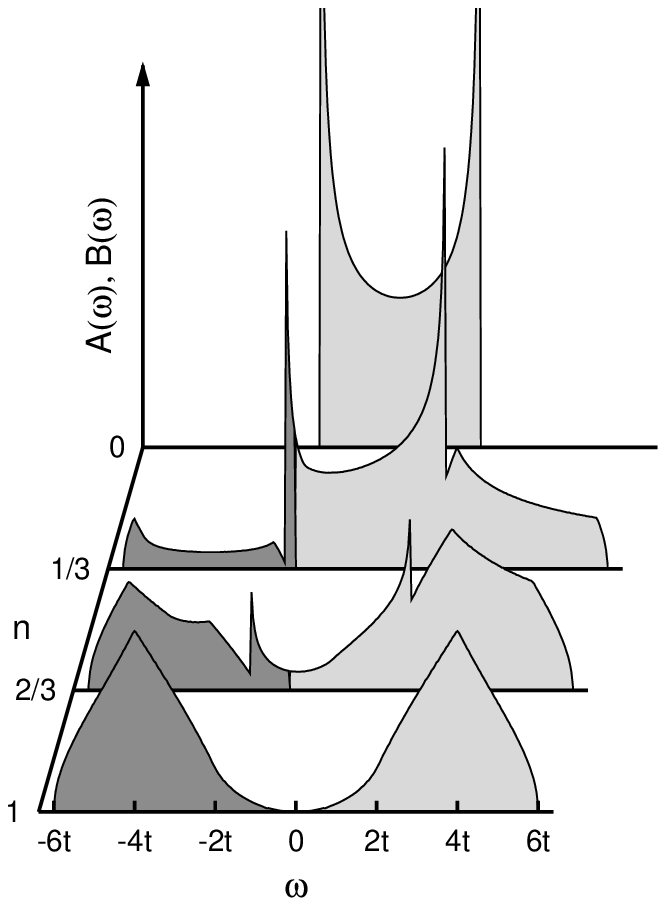}
\caption{
  The local spectral functions $B(\omega)$ (darker) and $A(\omega)$ 
(lighter
  shading) for $\alpha=\pi$. The filling increases from $n=0$ 
(top curve) to $n=1$ (bottom plot) in increments 
of $1/3$. 
 \label{fig:abwpi}}
\end{figure}

(ii) When  $\alpha=\pi$, the model actually corresponds to the 
electron-hole
symmetric correlated 
hopping model\cite{correlatedhopping} with $t_{AA}=t_{BB}=-t$ and 
$t_{AB}=t$ (the hopping amplitudes $t_{AA}$, $t_{BB}$ and $t_{AB}$ are defined
in Ref.~\onlinecite{correlatedhopping}):
\begin{equation}
  {\cal H} = -t \sum_{j,\sigma} 
  (1 \!-\! 2 \hat n_{j,\bar\sigma})
  (1 \!-\! 2 \hat n_{j+1,\bar\sigma})
  c^\dagger_{j,\sigma} c^{\phantom{\dagger}}_{j+1,\sigma} + {\rm H.c.} \,.
  \label{eq:hamcorrhop}
\end{equation}
The Hamiltonian (\ref{eq:hamcorrhop}) can be diagonalized
with the help of a unitary transformation
\begin{equation}
  \label{eq:ucorrhop}
  \tilde {\cal U} = \prod_{j=1}^L (-1)^{\hat n_{j,\uparrow} 
    \hat n_{j,\downarrow}} \,, 
\end{equation}
which is simpler than $\tilde {\cal U}=\tilde {\cal U}_1 \tilde {\cal U}_2$ 
given by Eqs.~(\ref{eq:U1}) and (\ref{eq:U2}),
 and it transforms the fermionic operators as
\begin{equation}
  \label{eq:fermicorrhop}
  \tilde {\cal U} c^\dagger_{j,\sigma} \tilde {\cal U}^\dagger 
  = 
  (1-2 \hat n_{j,\bar\sigma}) c^\dagger_{j,\sigma} \,, \quad
  \tilde {\cal U} c^{\phantom{\dagger}}_{j,\sigma} \tilde {\cal 
U}^\dagger = 
  (1-2 \hat n_{j,\bar\sigma}) c^{\phantom{\dagger}}_{j,\sigma} \,.
\end{equation}
so the transformed fermi operators remain ``local''. Furthermore, this
 transformation is not any more restricted to the 1D case.
 The evaluation of the matrix elements is now convenient for operators 
in
 site representation, and the matrix element in Eq.~(\ref{eq:akw}) 
becomes 
\begin{equation}
|\langle f | c^\dagger_{k,\uparrow} | {\rm GS} \rangle|^2 = L 
|\langle f | c^\dagger_{0,\uparrow} | {\rm GS} \rangle|^2  
\delta_{k,P_f-P_{\rm
    GS}} \,, \label{eq:mesite}
\end{equation}
where the $c^\dagger_{0,\uparrow}$ creates fermion on site $0$. Next, 
we 
 apply the canonical transformation to formulate the problem using
the transformed wave function [the analog of Eq.~(\ref{eq:canpsi})], 
and for 
the spectral function we get
\begin{equation}
  A_\uparrow(k,\omega) = L \sum_{\tilde f} 
  |\langle \tilde f | (1-2 \hat n_{j,\downarrow}) 
c^\dagger_{j,\uparrow} 
  | \widetilde {\rm GS} \rangle|^2 
 \delta(\omega-E_{f}^{N+1}+E_{\rm GS})
  \delta_{k,P_{f}^{N+1}-P_{\rm GS}} \,.
\nonumber
\end{equation}
The wave functions are product of the  spin-up and spin-down part, the
evaluation is straightforward and leads to
\begin{equation}
 A_\uparrow(k,\omega) = (1-2n_\downarrow)^2 A^{(0)}_\uparrow(k,\omega) 
+
   \frac{4}{L^2} 
 \sum_{q \in {\rm FS}_\downarrow}
 \sum_{q' \not\in {\rm FS}_\downarrow} 
 \sum_{k' \not\in {\rm FS}_\uparrow}  
\delta(\omega-\varepsilon_\downarrow(q')+\varepsilon_\downarrow(q)-\varepsilon_\uparrow(k'))
 \delta_{k,q'-q+k'} \,,
 \label{eq:ABpi}
\end{equation}
and a similar equation gives $B_\uparrow(k,\omega)$.  In the spectral function
we can identify the following two distinct features: 
(a) a Dirac-delta contribution
following the cosine-like dispersion, which is the reminder of the
noninteracting spectral function [Eq.~(\ref{eq:Akw0})] suppressed by a factor of
$(1-2n_\downarrow)^2$; (b) a broader continuum coming from the propagator
dressed with a single loop. 
  As we increase the filling, the weight
of the Fermi jump for zero magnetization ($n_\uparrow=n_\downarrow=n/2$) 
decreases as $(1-n)^2$, and will 
disappear at half filling,
leaving us with an $A(\omega) \propto \omega^2$ density of states 
[the $c_2\sim (1-n)^2$ in Eq.~(\ref{eq:aw_le}) for $\alpha =\pi$].
 To illustrate this behavior, we present the evolution of the
local spectral functions with density in Fig.~(\ref{fig:abwpi}).

(iii) General case: Like in the previous case, in evaluating the matrix
elements we use the site representation, given by Eq.~(\ref{eq:mesite}).
Next, we apply the canonical transformation to formulate the problem using the
transformed wave functions:
\begin{equation}
  \langle f | c^\dagger_{0,\uparrow} | {\rm GS} \rangle =
  \langle \tilde f | 
  c^\dagger_{0,\uparrow} e^{i \alpha \hat n_{0,\downarrow}}
  \hat R
  | \widetilde {\rm GS} \rangle e^{-i \alpha N_{\downarrow}} \,,
\end{equation}
where $\hat R = \prod_l e^{2 i \alpha l \hat n_{l,\downarrow}/L}$ 
[see Eq.~(\ref{eq:UU})].
As in the transformed basis the wave functions are product of the spin 
up and 
down free fermion wave functions,
$ | \widetilde {\rm GS} \rangle = | \widetilde {\rm GS}_\uparrow 
\rangle
| \widetilde {\rm GS}_\downarrow \rangle$ and 
$|\widetilde f \rangle = | \widetilde f_\uparrow \rangle
| \widetilde f_\downarrow \rangle$, 
the matrix element factorizes, and we get
\begin{eqnarray}
  A_\uparrow(k,\omega) &=& L \sum_{\tilde f} 
  |
  \langle \tilde f_\uparrow | c^\dagger_{0,\uparrow}  
  | \widetilde {\rm GS}_\uparrow \rangle  |^2 \times
  |\langle \tilde f_\downarrow | e^{i \alpha \hat n_{0,\downarrow}} 
  \hat R
  | \widetilde {\rm GS}_\downarrow \rangle 
|^2 
   \nonumber\\
 &&\times 
  \delta(\omega-E_{f,\uparrow}+E_{\rm GS,\uparrow}
     -E_{f,\downarrow}+E_{\rm GS,\downarrow})
  \delta_{k,P_{f,\uparrow}-P_{\rm GS,\uparrow}
            +P_{f,\downarrow}-P_{\rm GS,\downarrow}} \,.
\nonumber
\end{eqnarray}
In the equation above the $c^\dagger_{0,\uparrow}$ creates 
a fermion with energy $\varepsilon_\uparrow(k')$ and momentum 
$k' \not\in{\rm FS}_\uparrow$, in which case the matrix element is 
$| \langle \tilde f_\uparrow | c^\dagger_{0,\uparrow}  
  | \widetilde {\rm GS}_\uparrow \rangle  |^2=1/L $. This allows us to 
write the spectral function as a convolution
\begin{equation}
  A_\uparrow(k,\omega) = \frac{1}{L} \sum_{k' \not\in {\rm 
FS}_\uparrow} 
  A'_{\uparrow}(k-k',\omega-\varepsilon_\uparrow(k'))\,, \label{spectrumprimed}
\end{equation}
with
\begin{equation}
  A'_\uparrow(\omega,k) = L \sum_{\tilde f_\downarrow} 
  \left|   
    \langle \tilde f_\downarrow | e^{i \alpha \hat n_{0,\downarrow}} 
    \hat R
    | \widetilde {\rm GS}_\downarrow \rangle 
  \right|^2 
  \delta(\omega-E_{f,\downarrow}+E_{{\rm GS},\downarrow} )
  \delta_{k,P_{f,\downarrow}-P_{{\rm GS},\downarrow}}\,.
  \label{eq:Cdef}
\end{equation}
The interesting and nontrivial part of the calculation comes from the 
$\langle \tilde f_\downarrow | e^{i \alpha \hat n_{0,\downarrow}} 
\hat R
| \widetilde {\rm GS}_\downarrow \rangle$ matrix element. In the next 
and crucial step,
we eliminate the $e^{i \alpha \hat n_{0,\downarrow}}$. This can be 
easily 
accomplished  after the observation that 
translating the operator $\hat R$ a similar factor appears:
${\cal T} \hat R {\cal T}^\dagger = 
e^{2i \alpha (\hat n_{0,\downarrow}-n_\downarrow ) }  \hat R$.
So 
% \begin{eqnarray}
%  \langle \widetilde f_{\downarrow} |
%  \hat R
%  | \widetilde {\rm GS}_{\downarrow} \rangle 
%  \;=\; \langle \widetilde f_{\downarrow} | 
%  {\cal T}^\dagger {\cal T} \hat R
%  {\cal T}^\dagger {\cal T} | \widetilde {\rm GS}_{\downarrow} \rangle
%  \nonumber\\
%  \;=\;  e^{i (P_{f,\downarrow}-P_{{\rm GS},\downarrow}-2 \alpha n_\downarrow)}
%  \langle  \widetilde f_{\downarrow}
%  |  e^{2i \alpha \hat n_{0,\downarrow} } \hat R 
%  | \widetilde {\rm GS}_{\downarrow} \rangle
%\end{eqnarray}
%(we used that ${\cal T} c^\dagger_{k,\sigma} = 
%e^{-ik} c^\dagger_{k,\sigma}{\cal T}$). 
 \begin{equation}
  \langle  \widetilde f_{\downarrow}
  |  e^{2i \alpha \hat n_{0,\downarrow} } \hat R 
  | \widetilde {\rm GS}_{\downarrow} \rangle 
  = e^{i (2 \alpha n_\downarrow-P_{f,\downarrow}+P_{{\rm 
GS},\downarrow})}
  \langle \widetilde f_{\downarrow} |
  \hat R
  | \widetilde {\rm GS}_{\downarrow} \rangle \,.
\end{equation}
%(we used that ${\cal T} c^\dagger_{k,\sigma} = 
%e^{-ik} c^\dagger_{k,\sigma}{\cal T}$). 
Next, we note that 
$e^{i \alpha \hat n_{0,\downarrow}} = 
(e^{i \alpha} + e^{i 2 \alpha \hat n_{0,\downarrow}})/(1+e^{i 
\alpha})$, 
and we end up with
\begin{equation}
  \langle  \widetilde f_\downarrow
  | e^{i \alpha \hat n_{0,\downarrow} }
  \hat R
  | \widetilde {\rm GS}_\downarrow \rangle 
  =\frac{e^{i\alpha}+e^{i (P_{{\rm GS},\downarrow}-P_{f,\downarrow}
   +2 \alpha n_\downarrow)} }{1+e^{i\alpha}}
  \langle \tilde f_\downarrow | \hat R
  |\widetilde {\rm GS}_\downarrow \rangle \,. 
\end{equation}
To evaluate 
$ \langle \tilde f_\downarrow | \hat R |\widetilde {\rm GS}_\downarrow
\rangle$, we replace $|\widetilde {\rm GS}_\downarrow \rangle = \prod_j
c^\dagger_{k_j,\downarrow} |0 \rangle$ and $|\widetilde f_\downarrow 
\rangle 
= \prod_i c^\dagger_{k'_i,\downarrow}|0 \rangle $. Then we move 
$\hat R$ to the right across $c^\dagger_k$'s so that it acts to 
the 
vacuum, $\hat R|0 \rangle=|0 \rangle$. However, as 
$\hat R c^\dagger_{k,\downarrow}= c^\dagger_{k+(2\alpha/L),\downarrow} 
\hat R$
, the $k$ momenta are shifted by $2\alpha/L$ (this is equivalent to 
twisting
the boundary conditions):
\begin{equation} 
    \langle \tilde f_\downarrow | \hat R |\widetilde {\rm 
GS}_\downarrow
\rangle = \langle 0 |\prod_{i=1}^{N_\downarrow} c_{k'_{i},\downarrow} 
\hat R 
\prod_{j=1}^{N_\downarrow} c^\dagger_{k_j,\downarrow} |0\rangle 
  = \langle 0 |\prod_{i=1}^{N_\downarrow} c_{k'_{i},\downarrow} 
\prod_{j=1}^{N_\downarrow} c^\dagger_{k_j+\frac{2\alpha}{L},\downarrow}
 |0\rangle \,.
\end{equation}    
Here we have to calculate overlap of free fermion wave functions with 
different phase shifts due to the removal of a  $\uparrow$-spin 
fermion. This problem arises e.g. in the X-ray edge problem 
(Andersons's 
orthogonality catastrophe\cite{anderson67}), and the one-dimensional 
analog was discussed in Ref.~\onlinecite{penc97}. 
For the reader's convenience, we 
repeat here the main points. The anticommutation relation between the 
operators with different phase shifts reads
\begin{equation}
  A_{ij}=\left\{ 
  c^\dagger_{k_i+\frac{2\alpha}{L},\downarrow}, 
  c^{\phantom{\dagger}}_{k'_j,\downarrow} 
  \right\}
    = 
  \frac{e^{i\alpha}e^{\frac{i}{2}(k_i-k_j'+\frac{2\alpha}{L} )}}{L} 
  \frac{\sin \alpha}{\sin\left(\frac{k_i-k_j'}{2}+\frac{\alpha}{L} 
\right)}  \,.
\end{equation}
The overlap of the wave functions can be further calculated as 
$|\langle \tilde f_\downarrow | \hat R |\widetilde {\rm GS}_\downarrow
\rangle |^2 = |\det A_{ij}|^2$:
 \begin{eqnarray}
 |\langle \tilde f_\downarrow | \hat R |\widetilde {\rm GS}_\downarrow
 \rangle |^2
 &=& 
   \left|\left| 
     \begin{array}{ccc}
       \{c^\dagger_{k_1+2\alpha/L},c_{k'_1}\} & 
       \dots &
       \{c^\dagger_{k_1+2\alpha/L},c_{k'_{N_\downarrow}}\}  
     \\
       \vdots     &  \ddots     & \vdots      
     \\    
       \{c^\dagger_{k_{N_\downarrow}+2\alpha/L},c_{k'_1}\} & 
       \dots &
\{c^\dagger_{k_{N_\downarrow}+2\alpha/L},c_{k'_{N_\downarrow}}\}  
     \\
    \end{array}
    \right|\right|^2
   \nonumber\\
 &=& 
   \frac{\sin^{2N_\downarrow}\alpha}{L^{2N_\downarrow}} 
   \left|\left| \begin{array}{ccc}
     \displaystyle{
       \frac{1}{\sin \left( \frac{k_1-k'_1}{2}+\frac{\alpha}{L} 
\right)}
     } &
       \dots &
    \displaystyle{ 
       \frac{1}{\sin \left(
   \frac{k_1-k'_{N_\downarrow}}{2}+\frac{\alpha}{L} \right)} } 
     \\
       \vdots     & \ddots  & \vdots      
     \\
     \displaystyle{
       \frac{1}{\sin \left( 
\frac{k_{N_\downarrow}-k'_1}{2}+\frac{\alpha}{L} \right)}
     } &
       \dots &  
    \displaystyle{
       \frac{1}{\sin \left( 
\frac{k_{N_\downarrow}-k'_{N_\downarrow}}{2}+\frac{\alpha}{L} \right)}
     }
    \end{array}
    \right|\right|^2 \,.
   \nonumber
 \end{eqnarray}
 This determinant is actually a  Cauchy determinant and can be 
expressed as a product, so we end up with
\begin{equation}
|\langle \tilde f_\downarrow | \hat R |\widetilde {\rm GS}_\downarrow
\rangle|^2 
=  \frac{\sin^{2N_\downarrow} \alpha}
        {L^{2N_\downarrow}} 
  \frac{
    \prod_{j>i} \sin^2 \frac{k_j-k_i}{2}
    \prod_{j>i} \sin^2 \frac{k'_j-k'_i}{2}
  }{
    \prod_{j,i} \sin^2 \left(\frac{k'_i-k_j}{2}+\frac{\alpha}{L}\right)
 } \,. \label{eq:product}
\end{equation}
For the special $\alpha=0$ [where $A'_\uparrow(\omega,k) = L \delta(\omega)
\delta_{k,0}$] and $\alpha=\pi$ cases, taking the suitable limits, 
we recover the the results of
Eqs.~(\ref{eq:Akw0}) and (\ref{eq:ABpi}), respectively. In the $\alpha=\pi$
case the phase shift equals $2 \pi/L$, which is exactly the spacing between
two adjacent $k$ value. thus the orthogonality catastrophe is absent.

Following the same approach, for the photoemission part we get
\begin{equation}
   B_\uparrow(k,\omega) = \frac{1}{L} \sum_{k' \in {\rm FS}_\uparrow} 
         B'_{\uparrow}(\omega-\varepsilon_\uparrow(k'),k-k') \,,
\end{equation}
with
\begin{equation}
 B'_\uparrow(\omega,k) = L \sum_f 
 \left|   
 \langle \tilde f_\downarrow | e^{-i \alpha \hat n_{0,\downarrow}} 
\hat R^\dagger
 | \widetilde {\rm GS}_\downarrow \rangle 
 \right|^2 
  \delta(\omega-E_{{\rm GS},\downarrow}+E_{f,\downarrow} )
  \delta_{k,P_{{\rm GS},\downarrow}-P_{f,\downarrow}} \,.
\end{equation}

\begin{figure}
\includegraphics[width=8.5truecm]{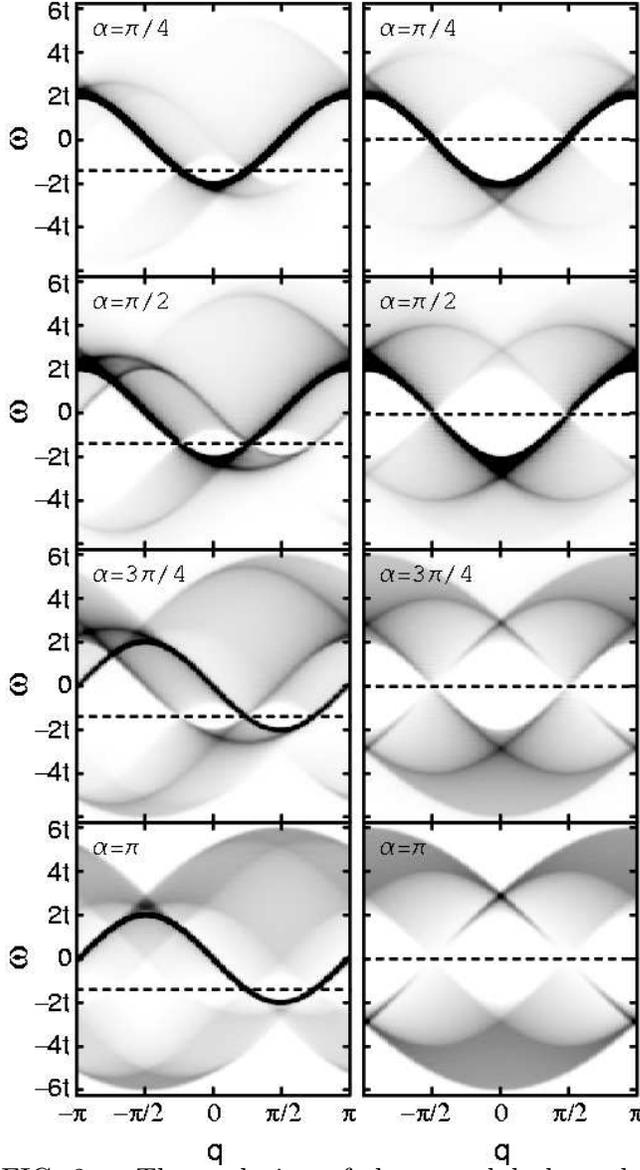}
\caption{
 The evolution of the $\omega$ and $k$ dependent spectral 
function as a
 function of $\alpha$ for $n=1/2$ (left) and $n=1$ (right plots). The shading
 is proportional to $A(k,\omega)$ and $B(k,\omega)$, the dashed line denotes
 the Fermi energy. The shift of the Fermi momenta
[Eq.~({\protect \ref{eq:knu}})] is compensated for by
 introducing  $q=k+\alpha n$ in the plot. 
We omitted the trivial 
$\alpha=0$ case.  
 \label{fig:abwkevol}}
\end{figure}

The product in Eq.~(\ref{eq:product}) can be evaluated numerically and 
spectral functions for relatively
large system can be obtained. 
The numerical result is presented in Fig.~\ref{fig:abwkevol} for some large
size systems. Starting from $\alpha=0$, we observe that there is an overall
shift in momentum proportional to $-2 \alpha n_{\downarrow}$ (which we
compensated for in the figure), and that apart
of the main contribution, which follows the cosine-like dispersion, additional
continuum-like features appear.  Finally, for even larger values of $\alpha$
another cosine-like feature appears with a considerable weight.

Alternatively, for the low energy part further analytical
considerations can be applied.\cite{penc97} Starting from 
Eq.~(\ref{eq:product}), 
 the weights of the peaks can be expressed via
$\Gamma$ functions in the $L\rightarrow \infty$ limit, 
leading to the power-law behavior of the
Luttinger liquid spectral function, and the exponents can be associated with
the phase shift.
We find singularities where the momenta of the final state are closely packed.
These happen at
\begin{equation}
  k^{(\nu)}_{\uparrow} = \nu \pi n_{\uparrow} - 2\alpha n_{\downarrow} 
\label{eq:knu}
\end{equation}
with $\nu$ an odd integer. The most important ones for small $\alpha$ are with
$\nu = \pm 1$, which coincides with the Fermi momenta $k^{\pm}_{F,\uparrow}$ .
As we can follow in Fig.~\ref{fig:abwkevol}, increasing $\alpha$ we get
weight for the tower at $k^{(3)}_{\uparrow}$, which eventually becomes
symmetric with $k^{(1)}_{\uparrow}$ for $\alpha=\pi$, while the weight of the
tower at $k^{(-1)}_{\uparrow}$ disappears at the same time. The behavior of
the primed spectral functions in Eq.~(\ref{spectrumprimed}) has 
a simple behavior near $k=0$:
\begin{equation}
  A'_\uparrow(k,\omega) \propto 
 \left[(\omega-\varepsilon_F)^2 
  - u^2 k^2\right]^{(\alpha/\pi)^2-1} \,, 
\end{equation}
while near $k=2 \pi n_\downarrow$:
\begin{equation}
   A'_\uparrow(k,\omega)  \propto
  \left[(\omega-\varepsilon_F)^2 
  - u^2(k-2 \pi n_\downarrow)^2\right]^{(\alpha/\pi-1)^2-1} \,. 
\end{equation}
This leads to the power law behavior of the  $A_\uparrow(k,\omega)$ as
presented in Eq.~(\ref{eq:powAkw}). The values of the exponents are tabulated
for some selected $\alpha$ in Table.~\ref{tbl:exp}.

The weight transfer can be quantified by observing the sum rules.
While the zeroth momentum is constant,
\begin{eqnarray}
  \int_{-\infty}^{\varepsilon_F} B_\uparrow(\omega) d \omega &=& 
n_\uparrow 
 \nonumber\\
  \int_{\varepsilon_F}^{+\infty} A_\uparrow(\omega) d \omega &=& 
1-n_\uparrow \,,
\end{eqnarray}
the first already shows the large weight transferred to energies far 
from the Fermi energy:
\begin{eqnarray}
  \int_{-\infty}^{\varepsilon_F} \omega B_\uparrow(\omega) d \omega &=& 
  \sum_i 
  \langle {\rm GS} |
  c^{\dagger}_{i,\uparrow} [H,c^{\phantom{\dagger}}_{i,\uparrow}] 
  | {\rm GS} \rangle
  \nonumber\\
  &=&
  - \frac{2 t}{\pi} \sin(\pi n_\uparrow) 
  - \frac{4t}{\pi} n_\uparrow \sin(\pi n_\downarrow) (1-\cos\alpha),
 \nonumber\\
  \int_{\varepsilon_F}^{+\infty} \omega A_\uparrow(\omega) d \omega &=&
  \sum_i 
  \langle {\rm GS} |
  c^{\phantom{\dagger}}_{i,\uparrow} [H,c^\dagger_{i,\uparrow}] 
  | {\rm GS} \rangle
  \nonumber\\
  &=&
 \frac{2 t}{\pi} \sin(\pi n_\uparrow) 
 + \frac{4t}{\pi}(1-n_\uparrow) \sin(\pi n_\downarrow) (1-\cos\alpha) \,.
\end{eqnarray}
We can see that the weight transfer to higher energies is the largest for
$\alpha=\pi$ and at half-filling. 

\section{Conclusions}

We have presented the exact one electron Green's function for a model fermi
system in 1D with a non Fermi liquid behavior for essentially any value of
the interaction strength.  A few exact analytical calculations of the spectral
function for model systems, such as the $1/r^2$ exchange t-J model, with a
projection to single occupancy are available in
literature.\cite{arikawa,kato}  The Green's function for this system
obtained here does require some numerics, and is not totally analytical.
However unlike the situation in projected models, such as the t-J model , it
satisfies the sum rules familiar from text books for weakly interacting Fermi
liquids (e.g. the complete electron sum rule with large $\omega$ behavior of
$G$ as 1/ $\omega$).  This feature makes the present model particularly
interesting in the context of the programme of reconstruction of the spectral
function from its moments (e.g. see Ref.~\onlinecite{arpes}).

\acknowledgements
K. P. acknowledges the support of the Hungarian OTKA D32689 and Bolyai 118/99.

\begin{figure}
\includegraphics[width=8.5truecm]{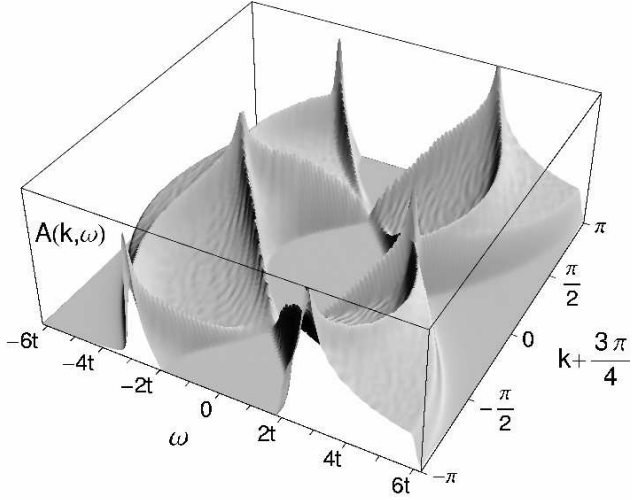}
\caption{
 The spectral function for $\alpha=3\pi/4$ and $n=1$. Here the Fermi energy is
 at $\omega=0$.
 \label{fig:abwkevol2}}
\end{figure}

\begin{figure}
\includegraphics[width=7.0truecm]{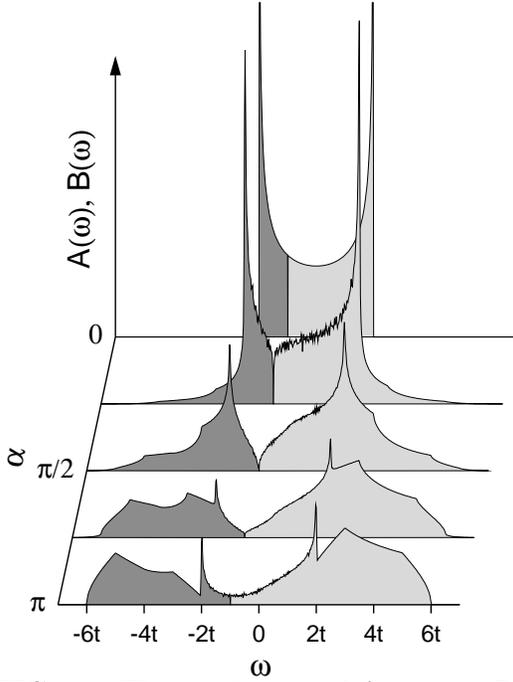}
\caption{
  The local spectral functions $B(\omega)$ (darker) and $A(\omega)$ 
(lighter
  shading) for $n=2/3$. The $\alpha$ changes from $0$ 
(noninteracting case, top curve) to $\pi$ (bottom plot) in increments 
of
  $\pi/4$. To minimize finite size effects, the curves show the average
of $L=303$, 279, 255, 231, 207 and 183.  
 \label{fig:abw}}
\end{figure}

\begin{figure}
\includegraphics[width=7.5truecm]{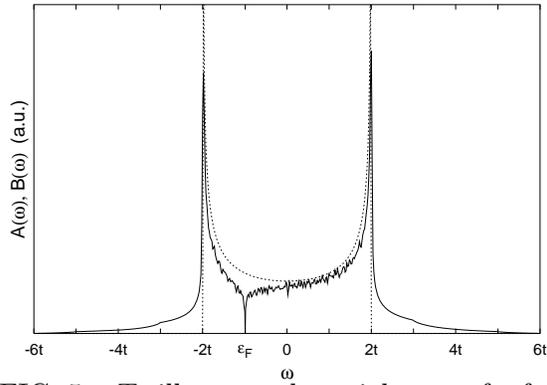}
\caption{
  To illustrate the weight transfer for small $\alpha$, we compare 
the local spectral function for $\alpha=\pi/4$ (solid line) to the 
$\alpha=0$ case (dashed). The $\alpha=\pi/4$ case behaves as 
$A(\omega) \sim |\omega-\varepsilon_F|^{1/8}$ near the Fermi energy.
 \label{fig:abw1o4}}
\end{figure}

\begin{table}
\begin{tabular}{cccccc}
  $\alpha$           & 0 & $\pi/4$ & $\pi/2$ & $3\pi/4$ & $\pi$\\
\hline
 $2(\alpha/\pi)^2$   & 0 & 1/8     & 1/2     & 9/8      & 2     \\
 $2(\alpha/\pi-1)^2$ & 2 & 9/8     & 1/2     & 1/8      & 0    \\
\end{tabular}
\caption{The exponents in the local spectral function 
[Eq.~({\protect \ref{eq:aw_le}})]. 
\label{tbl:exp}}
\end{table}

\end{document}